\documentclass[prl,aps,floatfix,twocolumn,superscriptaddress,longbibliography,10pt]{revtex4-2}

\usepackage{amsmath,amssymb}
\usepackage{graphicx}
\usepackage[dvipsnames]{xcolor}
\usepackage[colorlinks,linkcolor=black,citecolor=black,urlcolor=black,filecolor=black]{hyperref}
\usepackage{amsthm}
\usepackage{enumerate}
\usepackage{setspace}
\usepackage{threeparttablex}
\usepackage[export]{adjustbox}
\renewcommand\thesection{\arabic{section}}
\DeclareMathAlphabet{\mymathbb}{U}{BOONDOX-ds}{m}{n}
\usepackage{dsfont}
\usepackage{braket}
\hypersetup{
    pdfstartview={FitH}, 
    colorlinks=true, 
    linkcolor=blue, 
    citecolor=blue, 
    filecolor=magenta,
    urlcolor=blue
}
\usepackage{siunitx}

\usepackage{bm}
\usepackage{mathtools}
\usepackage{url}
\usepackage{enumitem}
\usepackage[capitalise]{cleveref}
\usepackage{verbatim}
\usepackage{MnSymbol}
\usepackage{physics}
\usepackage{stmaryrd}

\usepackage{pifont}
\newcommand{\cmark}{\ding{51}}%
\newcommand{\xmark}{\ding{55}}%
\usepackage{booktabs}
\usepackage{makecell}
\AtBeginDocument{%
  \heavyrulewidth=.08em
  \lightrulewidth=.05em
  \cmidrulewidth=.03em
  \belowrulesep=.65ex
  \belowbottomsep=0pt
  \aboverulesep=.4ex
  \abovetopsep=0pt
  \cmidrulesep=\doublerulesep
  \cmidrulekern=.5em
  \defaultaddspace=.5em
}

\begin{document}

\title{Constant-Overhead Fault-Tolerant Bell-Pair Distillation using High-Rate Codes}

\author{J. Pablo Bonilla~Ataides}
\email{jbonillaataides@g.harvard.edu}
\affiliation{Department~of~Physics,~Harvard~University,~Cambridge,~MA~02138,~USA}

\author{Hengyun~Zhou}
\affiliation{QuEra Computing Inc., Boston, MA 02135, USA}

\author{Qian~Xu}
\affiliation{Institute for Quantum Information and Matter, Caltech, Pasadena, CA, USA}
\affiliation{Walter Burke Institute for Theoretical Physics, Caltech, Pasadena, CA, USA}

\author{Gefen~Baranes}
\affiliation{Department~of~Physics,~Harvard~University,~Cambridge,~MA~02138,~USA}
\affiliation{Department of Physics and Research Laboratory of Electronics, Massachusetts Institute of Technology, Cambridge, MA, USA}

\author{Bikun~Li}
\affiliation{Pritzker School of Molecular Engineering, The University of Chicago, Chicago, Illinois 60637, USA}

\author{Mikhail~D.~Lukin}
\email{lukin@physics.harvard.edu}
\affiliation{Department~of~Physics,~Harvard~University,~Cambridge,~MA~02138,~USA}

\author{Liang~Jiang}
\email{liang.jiang@uchicago.edu}
\affiliation{Pritzker School of Molecular Engineering, The University of Chicago, Chicago, Illinois 60637, USA}


\begin{abstract}
We present a fault-tolerant Bell-pair distillation scheme achieving constant overhead through high-rate quantum low-density parity-check (qLDPC) codes.
Our approach maintains a constant distillation rate equal to the code rate while requiring no additional overhead beyond the physical qubits of the code.
Full circuit-level analysis demonstrates fault-tolerance for input Bell pair infidelities below a threshold $\sim 10\%$, readily achievable with near-term capabilities.
Unlike previous proposals, our scheme keeps the output Bell pairs encoded in qLDPC codes at each node, eliminating un-encoding overhead and enabling direct use in distributed quantum applications through recent advances in qLDPC computation.
These results establish qLDPC-based distillation as a practical route toward resource-efficient quantum networks and distributed quantum computing.
\end{abstract}

\maketitle

\begin{figure*}[ht]
    \centering
    \includegraphics[width=1.0\textwidth]{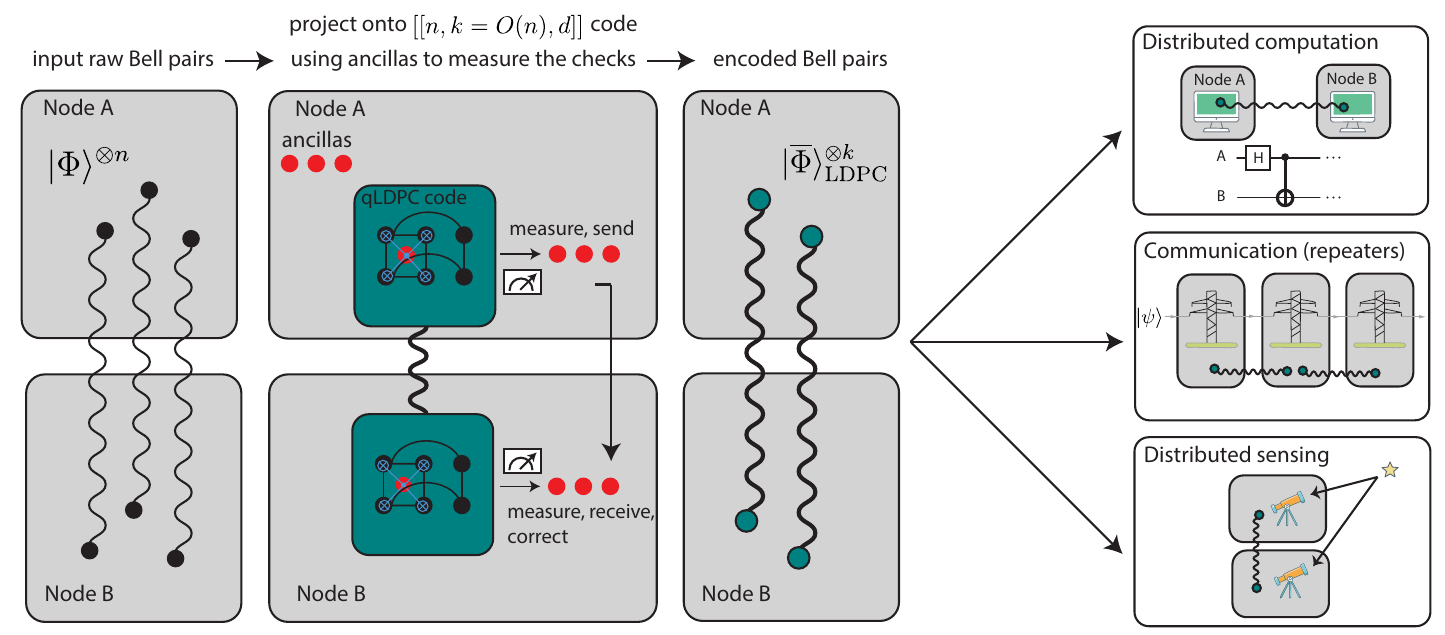}
    \caption{Constant-overhead Bell-pair distillation using qLDPC codes.
    Alice and Bob share $n$ initial raw Bell pairs.
    They measure the checks of a qLDPC code on each side using local ancilla qubits.
    Alice communicates her measurement results to Bob, who receives them and performs error correction, projecting the Bell pairs into the code state. 
    This process results in a constant distillation rate equal to the rate of the qLDPC code, requiring only one-way communication.
    By performing logical operations on qLDPC codes, the encoded Bell pairs can be used for various applications, including distributed computation, communication via repeaters, and distributed sensing.}
    \label{fig:encoding}
\end{figure*}

As quantum systems scale, establishing high-fidelity connections between nodes will be essential for distributed quantum computation, communication, and sensing~\cite{monroe2014large,muralidharan2016optimal,ge2018distributed,proctor2018multiparameter,zhuang2018distributed,eldredge2018optimal,yang2024asynchronous} (see Fig.~\ref{fig:encoding}).
A promising approach to achieving this is through shared Bell pairs, where one qubit of each pair is located at each node.
Bell pairs provide non-local entanglement, enabling quantum information transfer through local operations and classical communication.

However, raw Bell pairs generated by current hardware are too noisy for practical applications.
Experiments can produce Bell pairs with infidelities of $ \sim 5\%$~\cite{stephenson2020high,mirhosseini2020superconducting,jing2019entanglement,knaut2024entanglement,meesala2024quantum}, whereas distributed algorithms may require error rates below $10^{-10}$.
To bridge this gap, entanglement distillation -- also known as  purification -- is used to convert multiple noisy Bell pairs into fewer, higher-fidelity pairs~\cite{bennett1996purification}.  

A practical Bell-pair distillation scheme should satisfy several key criteria.  
It must be scalable, which can be achieved by maintaining constant overhead.
It should also be fault-tolerant, ensuring robustness against both network errors and local gate errors.
Additionally, it should exhibit error thresholds for Bell-pair infidelity and local gate errors that are both feasible and achievable with near-term hardware.
Finally, the scheme should minimize additional resource requirements, such as classical communication costs, code decoding complexity, and memory overhead from post-selection in non-deterministic schemes.

Topological codes are among the leading proposals for fault-tolerant quantum computing, offering high thresholds, local low-weight checks, and compatibility across multiple architectures~\cite{kitaev2003fault,fowler2012surface,bonilla2021xzzx,egan2021fault,google2023suppressing,bluvstein2023logical,google2024quantum}.
However, these codes encode only a constant number of logical qubits, resulting in a vanishing asymptotic rate.
As a result, distillation schemes relying on topological codes, such as lattice surgery schemes~\cite{litinski2019game,fowler2010surface,ramette2023fault,sinclair2024faulttolerant,leone2024upper}, require a large number of physical Bell pairs to achieve the low error rates demanded by logical algorithms.

\begin{table*}[ht]
\centering
\renewcommand{\arraystretch}{1.3}
\resizebox{\textwidth}{!}{%
\begin{tabular}{@{}lccccc@{}}
\toprule
\textbf{Method} & \textbf{Overhead} & \makecell{\textbf{Classical} \\ \textbf{Communication}} & \makecell{\textbf{Local Gate} \\ \textbf{Error Tolerance}} & \textbf{Encoding Resources} & \textbf{Threshold} \\ 
\midrule
BDSW-2EPP~\cite{bennett1996mixed} & $> O(1)$ & Two-way & \xmark & Non-deterministic, requires buffer memory & \makecell{50\% with perfect \\ local ops.} \\
BDSW-1EPP~\cite{bennett1996mixed} & $O(1)$ & One-way & \xmark & Requires decoding random quantum codes & Not practical \\
Lattice surgery~\cite{fowler2010surface,ramette2023fault,sinclair2024faulttolerant,leone2024upper} & $> O(1)$ & One-way & \cmark & $O(d)$ time$^\ddagger$ & $\sim 10\%$ \\
Shi et al.~\cite{shi2024stabilizer} & $O(1)$ & Two-way & \xmark & $O(1)$ & Not studied \\
Pattison et al.~\cite{pattison2024fast} & $O(1)$ & Two-way & \cmark & Non-deterministic, requires buffer memory & \makecell{50\% with perfect \\ local ops.$^\dagger$} \\
\textbf{This work} & $O(1)$ & One-way & \cmark & $O(1)$ & \makecell{$\sim 10 \%$ full circuit\\-noise simulation} \\
\bottomrule
\end{tabular}%
}
\caption{Summary of selected Bell-pair distillation methods.
\textbf{Overhead} refers to the asymptotic number of input physical Bell pairs required per output logical Bell pair, assuming perfect local operations.
\textbf{Classical communication} indicates whether the method relies on error detection (requiring two-way classical communication) or error correction (necessitating only one-way communication).
\textbf{Local gate error tolerance} specifies the protocol's robustness against local gate errors.
Protocols that directly un-encode to physical Bell pairs are not robust against local gate errors.
\textbf{Encoding resources} refer to any additional resources required by the protocol.
\textbf{Threshold} represents the scheme's threshold against network errors.
Our scheme is simulated at the full circuit-noise level, assuming a local gate error rate of $0.1\%$.
For further details on each method, refer to the Supplementary Information~\cite{SM}.
\\ $^\dagger$: Imperfect local operations reduce performance only slightly.
\\ $^\ddagger$: Although the primary scheme analyzed in these works involves lattice surgery at the surface code boundary, Ref.~\cite{ramette2023fault} suggests that the results extend to transversal gates, where the time cost can be reduced to $O(1)$.
}
\label{tab:methods}
\end{table*}

Quantum low-density parity-check (qLDPC) codes provide a promising approach for constant-rate distillation~\cite{breuckmann2021quantum,shi2024stabilizer,rengaswamy2024entanglement,sutcliffe2025distributed,shi2025measurementbased,menon2025magic}, but existing proposals lack a full fault-tolerant analysis and typically un-encode to physical Bell pairs, limiting robustness to local gate errors.
To the best of our knowledge, no qLDPC distillation proposal has yet been analyzed at the full circuit-noise fault-tolerance level.
Ref.~\cite{shi2024stabilizer} suggests a fault-tolerant encoder for qLDPC codes, similar to the approach in this work, but does not study specific codes or conduct threshold simulations. 
The use of qLDPC codes for state distillation has also been proposed in Ref.~\cite{rengaswamy2024entanglement}, though in a different context; to distill GHZ states for distributed quatum error correction (QEC).
Similarly, Ref.~\cite{sutcliffe2025distributed} explores hyperbolic Floquet codes, implementing them non-locally by distributing qubits across multiple nodes.
In contrast, we consider a different approach, assuming local nodes large enough to implement QEC internally~\cite{bluvstein2023logical,google2024quantum}. 

An alternative scheme for achieving constant-rate distillation relies on QEC codes for error detection, where errors are flagged and discarded instead of corrected.
Recent work has demonstrated constant-rate distillation using code concatenation~\cite{pattison2024fast,yamasaki2024time} and via scrambling~\cite{gu2025constant}.
However, these error-detection schemes are non-deterministic and require two-way classical communication for post-selection.
Additionally, this approach may require additional buffer memory overhead, otherwise, faulty pairs will bottleneck the execution by requiring one to wait until the next distillation round completes.

In this Letter, we present a scheme that satisfies all the key requirements for practical, constant-overhead entanglement distillation.
We compare our method with other leading distillation schemes across multiple performance metrics in Table~\ref{tab:methods}.
Our approach utilizes constant-rate qLDPC codes, following a stabilizer protocol~\cite{brun2006correcting,wilde2010convolutional}, where code checks are measured to project Bell pairs onto the encoded code state (see Fig.~\ref{fig:encoding}).

We consider codes that are feasible for implementation in reconfigurable atom arrays~\cite{xu2024constant,bluvstein2022quantum,bluvstein2023logical,evered2023high,bluvstein2025architectural,bonillaataides2025neural} and possess high rates.
We perform a full fault-tolerant circuit-level analysis of our scheme and observe high thresholds achievable with near-term hardware~\cite{li2024high}.
Furthermore, recent advances in qLDPC gates enable us to leave the output Bell pairs encoded in the code.
This approach eliminates the extra un-encoding stage required in many current distillation schemes, allowing us to achieve distillation fidelities that are not limited by gate errors.

\textit{Ancilla-assisted encoding.---} We now describe the protocol used to distill Bell pairs using constant-rate qLDPC codes, as illustrated in Fig.~\ref{fig:encoding}.
The goal of entanglement distillation is to take noisy input Bell pairs, $\ket{\Phi^+}^{\otimes n}$, and output higher-fidelity Bell pairs, $\ket{\bar{\Phi}^+}^{\otimes k}$, where $\ket{\Phi^+} = (\ket{00}+\ket{11})/\sqrt{2}$ and $k \leq n$.

This process can be associated with a QEC code, where redundancy is used to achieve lower error rates~\cite{bennett1996mixed,bennett1996purification}.
A quantum code is labeled as $[[n,k,d]]$, where $n$ is the number of physical qubits, $k$ is the number of encoded logical qubits and $d$ is the code's distance.
A code with distance $d$ can detect up to $d-1$ errors and correct up to $\lfloor \frac{d-1}{2} \rfloor$ errors.
The code rate, defined as $R = k/n$, quantifies the encoding efficiency of the code.
If the QEC scheme is fault-tolerant, there exists a threshold physical error rate below which increasing the code distance can exponentially suppress the error rate of the encoded logical qubits, enabling arbitrarily low error rates.

A stabilizer quantum code is specified by stabilizer generators $\mathcal{S} = \langle g_1, \dots, g_m\rangle $, where each $g_i \in \mathcal{P}^n$ is an $n$-qubit Pauli operator.
The codespace of the code is the simultaneous $+1$ eigenspace of the stabilizers of the code, satisfying $g \ket{\bar{\psi}} = (+1) \ket{\bar{\psi}}$ for all $g \in \mathcal{S}$.
We focus on CSS codes, whose stabilizer generators are divided into $X$-type ($g_x\in \{X,I\}^{\otimes n}$) and $Z$-type ($g_z\in\{Z,I\}^{\otimes n}$).

Quantum low-density parity-check (qLDPC) codes are those in which each check is supported on a constant number of qubits, and each qubit participates in a constant number of checks. 
This bounded check and qubit degree ensures that the syndrome extraction circuit remains constant-depth, facilitating the fault-tolerance of the QEC protocol.
When qLDPC codes are used to distill Bell pairs, the distillation rate is constant and equal to the code rate.
Although the checks of qLDPC codes can be highly non-local~\cite{baspin2022connectivity,baspin2022quantifying,baspin2024improved}, various schemes and architectures have been developed to implement them efficiently~\cite{xu2024constant,bravyi2024high,stein2024architectures,viszlai2023matching,poole2025architecture}.

To perform error correction, the stabilizers (or checks) are measured and a decoding algorithm is used to infer the error corresponding to any $-1$ measurement outcomes.
These $-1$ outcomes collectively form the syndrome of the code.
A single ancilla per check can be used to measure the value of each check by entangling the ancilla with the data qubits involved in the check and then measuring the ancilla.
Single-ancilla syndrome extraction circuits can be made fault-tolerant due to the bounded weight of the checks and further can be made distance preserving for some code families~\cite{tremblay2022constant,manes2023distancepreserving,xu2024constant}.

By measuring the stabilizers of the code on the qubits of the Bell pairs at each node, Alice and Bob can project the Bell pairs into the codespace.
Specifically, Alice uses local ancilla qubits to measure the checks of the qLDPC code on her side and sends the syndrome to Bob through a classical channel.
Bob then measures the checks of the code on his side and adds his measurements to the syndrome received from Alice.
After Bob performs error correction on the joint checks, the resultant state consists of $k$ encoded Bell pairs in the codespace of the qLDPC code.
A rigorous analysis of this procedure, along with a proof of the protocol’s fault tolerance, is provided in the End Matter.

\begin{figure*}[ht]
    \centering
    \includegraphics[width=\textwidth]{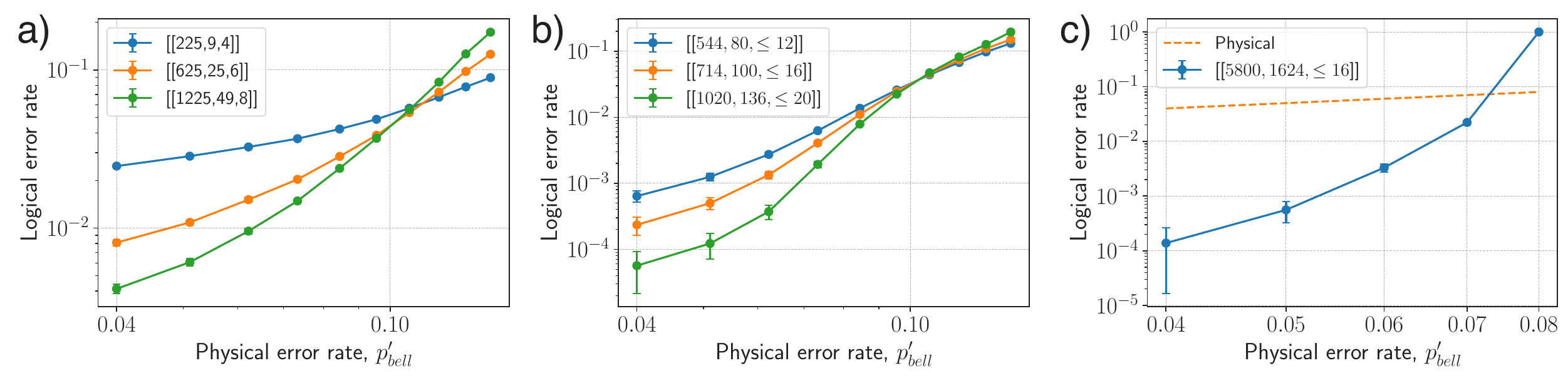}
    \caption{Fault-tolerant simulation of qLDPC distillation.
    The Bell pairs experience input depolarizing noise with effective strength $p'_{\text{bell}}$.
    The circuit undergoes two-qubit gate depolarizing errors at a rate of $p' = 2 \times 0.1\%$.
    We simulate 14 decoding rounds with 3 cycles of syndrome extraction per round (42 cycles in total) and decode each round with BP, using BP+OSD for the final round.
    (a), (b), (c) Numerical results for the HGP, LP, and SC codes, respectively.
    We find a threshold $p_{\text{bell}}' \sim 10\%$ for the HGP and LP code families and a pseudothreshold of $p_{\text{bell}}\sim 7.5\%$ for the SC code.
    We calculate the logical error rate for the entire block per cycle, using $\bar{P} = 1 - (1-\bar{P}_{\text{tot}})^{1/42}$, where $\bar{P}_{\text{tot}}$ is the total error after 42 cycles.
    Error bars are calculated assuming a binomial distribution, with $N$ shots, giving $\sqrt{P_{\text{tot}} (1- P_{\text{tot}})/N}$.
    }
    \label{fig:results}
\end{figure*}

\textit{Error modeling.---}
We perform a full fault-tolerant analysis by simulating the encoding of Bell pairs into qLDPC codes, incorporating both Bell-pair and local gate errors -- extending beyond prior studies that typically assume only depolarizing noise on input pairs.

Each Bell pair undergoes a depolarizing channel with strength $p_{\text{bell}}$, while local two-qubit gates in the error correction circuit experience depolarizing noise with strength $p$.
To simplify the simulation, we leverage symmetry in Bell-pair errors to shift all noise onto Bob's side, leaving Alice's side noiseless.
This approximation allows us to simulate only Bob’s subsystem while preserving the overall logical error behavior.
In the End Matter and Supplementary Information~\cite{SM}, we show that the increased noise on Bob's side can be approximated as $p'_{\text{bell}} \approx 2p_{\text{bell}}$ and $p' \approx 2p$.
Following convention in prior work~\cite{ramette2023fault,rengaswamy2024entanglement,pattison2024fast,gu2025constant,shi2025measurementbased}, we quote \( p_{\text{bell}}' \) as the input Bell error rate.
This convention aligns with experimental setups where Alice generates the Bell pairs at her node, and only the qubit sent to Bob experiences the noisy fiber channel.

\textit{Numerical simulations.---}
We perform full fault-tolerant circuit-noise simulations using three different classes of qLDPC codes: hypergraph product (HGP) codes with a rate of $\sim 4\%$~\cite{tillich2014quantum}, quasi-cyclic lifted product (LP) codes with a rate of $\sim11\%$~\cite{raveendran2022finite}, and a spatially coupled (SC) code with a rate of $\sim 1/3$~\cite{yang2023spatially,hagiwara2011spatially}.
To the best of our knowledge, a distillation rate of 1/3 is the highest rate achieved by any deterministic and efficiently decodable distillation protocol to date~\footnote{The quoted rates for the LP and SC codes correspond to finite-size instances.
For further details, see the End Matter and Supplementary Information~\cite{SM}.}.
Further details on code constructions can be found in the End Matter.

The fault-tolerant simulations include multiple rounds of the syndrome extraction circuit, followed by a final transversal measurement of the data qubits of the code.
To extract the values of the qLDPC checks, we use a single ancilla qubit, which is entangled with the data qubits in the support of each check.
A measurement of the ancilla qubit then provides the syndrome information used for decoding.
Fault tolerance is maintained by scheduling the entangling gates between the ancilla and data qubits in parallel, following the coloration circuit approach~\cite{tremblay2022constant,xu2024constant}.
We note that accounting for movement errors required to implement non-local checks on a reconfigurable atom array platform introduces negligible additional errors, as idling error is low and compact rearrangement schemes have been developed~\cite{xu2024constant,SM}.

To decode, we turn the circuit into a spacetime code and use the BP+OSD decoding algorithm~\cite{panteleev2019degenerate, roffe2020decoding, higgott2023improved, xu2024constant}.
The checks of the spacetime code correspond to products of stabilizers at subsequent time steps, and the ``qubits" of the spacetime code represent fault locations in the circuit.
We leverage the single-shot property of the codes, which requires only $O(1)$ cycles of QEC per decoding round to achieve fault tolerance against measurement errors~\cite{quintavalle2020single, xu2024constant}.
Although, in principle, a single QEC cycle per round is sufficient for fault tolerance, we implement three cycles per round to provide additional redundancy against measurement errors.
Our simulations consist of 14 rounds of 3 cycles each, totaling 42 cycles.
We decode each round with BP only, and to project back into the code space, we decode the final round with BP+OSD.
The circuits are constructed in Stim~\cite{gidney2021stim}, and we use the BP+OSD decoder implementation provided in the ldpc library~\cite{Roffe_LDPC_Python_tools_2022}.

For the error model, we assume local gate errors with strength $p = 0.1\%$, leading to an effective error strength on Bob's side of $p' = 2 \times 0.1\%$ for the two-qubit gates in the syndrome extraction circuits.
We vary the input Bell error $p_{\text{bell}}'$ and calculate the logical error rate of the qLDPC block per cycle.
The results are presented in Fig.~\ref{fig:results}.
We observe a threshold of $p_{\text{bell}}'\sim 10\%$ for the HGP and LP code families and a pseudothreshold of $p_{\text{bell}}\sim 7.5\%$ for the SC code.
Since we plot the block error rate, meaning that a failure is declared if any of the logical qubits of the qLDPC code fails, the error rates per logical qubit can be substantially lower than the values shown in Fig.~\ref{fig:results}.

\begin{figure*}[h]
    \centering
    \includegraphics[width=\textwidth]{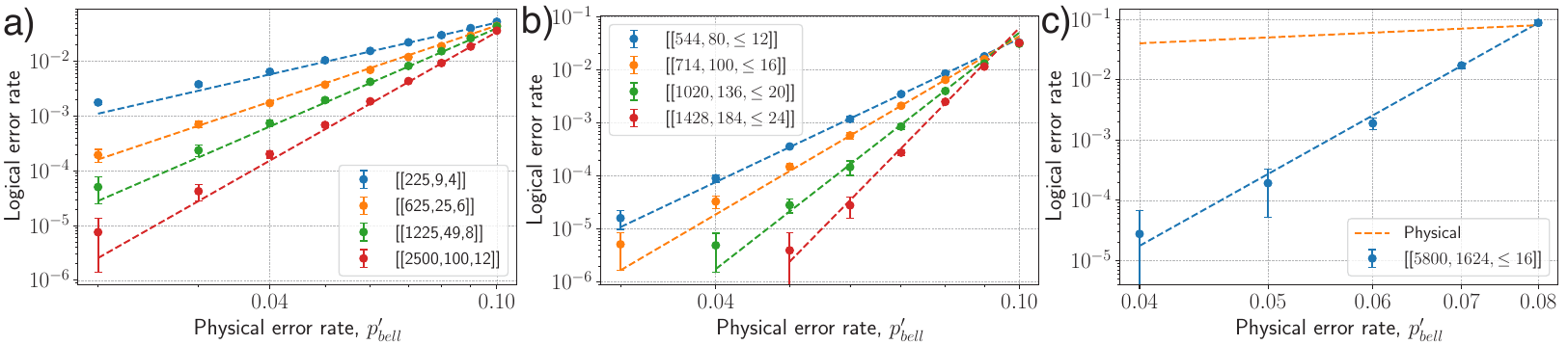}
    \caption{Sub-threshold data and fitted ansatz.
    Bell pairs have input depolarizing noise with strength \( p_{\text{bell}}' = p \), and two-qubit gate depolarizing errors at \( p_{\text{2q}}' = p/50 \).
    We simulate 14 decoding rounds, each with 3 syndrome extraction cycles (42 total).
    Each round is decoded using BP, with the final round using BP+OSD.
    (a) Numerical results for the HGP codes.
    (b) Numerical results for the LP codes.
    (c) Numerical results for the SC code.
    Dotted lines fit the sub-threshold data to the ansatz \( \bar{P} = A(p/p_{\text{th}})^{Bn^C} \), where \( n \) is the code size, \( p_{\text{th}} \) is the threshold error rate, and \( A, B, C \) are fitting parameters.
    The logical error rate per cycle is \( \bar{P} = 1 - (1-\bar{P}_{\text{tot}})^{1/42} \), where \( \bar{P}_{\text{tot}} \) is the total error after 42 cycles.
    Error bars assume a binomial distribution with \( N \) shots, computed as \( \sqrt{P_{\text{tot}} (1 - P_{\text{tot}})/N} \).
    }
    \label{fig:subthr}
\end{figure*}

The plateau observed in the logical error rate at low values of \( p_{\text{bell}}' \) arises from the use of a fixed circuit-level gate error.
If we instead allow the local gate noise to vary along with the Bell error rate, this plateau disappears, enabling us to study the subthreshold scaling behavior of the codes.
In the End Matter, Fig.~\ref{fig:subthr} and Table~\ref{tab:ler}, we examine the subthreshold scaling of the codes and tabulate the achievable error rates for various instances.
For example, we find, by extrapolation, that the SC code can achieve a logical error rate of $\sim 2\times 10^{-12}$ at input physical error rate of $1\%$.

\textit{Outlook.---}
We presented and analyzed a one-way constant-rate distillation scheme based on qLDPC error correction that achieves high fault-tolerant thresholds, rates as high as 1/3, and requires no additional overhead beyond the qubits of the qLDPC code.
Leaving the Bell pairs encoded at each node eliminates the unencoding step required in many existing schemes and ensures protection against local errors for future computations.

However, performing local operations on these encoded Bell pairs requires the ability to manipulate qLDPC-encoded qubits.
In this direction, numerous resource-efficient schemes have recently been proposed to enable selective computation with qLDPC codes~\cite{xu2024fast,cross2024linear,williamson2024lowoverhead,breuckmann2024fold,swaroop2024universal,breuckmann2024cups,he2025quantum,alex2025computing}. 

\begin{figure}[ht]
    \centering
    \includegraphics[width=\columnwidth]{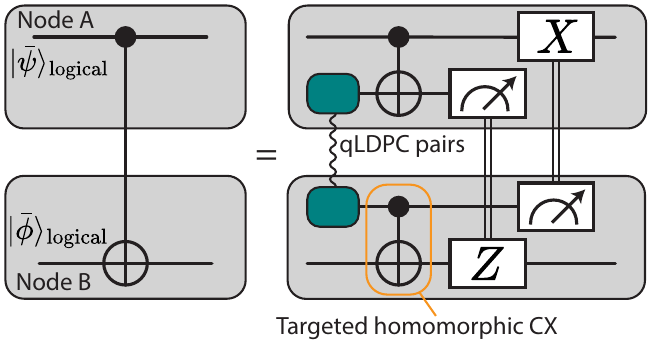}
    \caption{Distributed CNOT using qLDPC-distilled Bell pairs.
    The distributed CNOT gate can be implemented using local targeted CNOT operations between logical qubits and one of the \( k \) encoded Bell pairs of the qLDPC code.
    These targeted CNOT gates can be performed using lattice surgery techniques or homomorphic gadgets.
    In product codes such as HGP codes, transversal approaches using homomorphic gadgets provide a constant-overhead implementation compatible with reconfigurable neutral atom architectures.
    }
    \label{fig:distributed-cx}
\end{figure}

As an example, consider the implementation of a distributed CNOT gate, as illustrated in Fig.~\ref{fig:distributed-cx}.
This gate can be realized using the high-fidelity Bell pairs distilled with the qLDPC code and local logical CNOT gates.
The local logical CNOTs must target a single Bell pair among the $k$ encoded pairs in the qLDPC code.

This targeted CNOT operation can be implemented in various ways.
Lattice surgery techniques~\cite{cohen2022low,xu2024constant,williamson2024lowoverhead,ide2024faulttolerant,swaroop2024universal,cross2024linear,cross2024improved} provide a viable approach, while certain product codes, such as HGP codes, enable a more efficient implementation via homomorphic gadgets~\cite{xu2024fast}.
The latter maintains constant overhead and is naturally compatible with the movement capabilities of reconfigurable atom arrays~\cite{xu2024constant,xu2024fast}.
In the proposed implementation, the error rate of the distributed CNOT gate is primarily determined by the error rate of the qLDPC-encoded Bell pairs, as this is the only stage where nonlocal (and noisy) physical elements are involved.
Thus, the results of Fig.~\ref{fig:results} are also applicable to the distributed CNOT gadget shown in Fig.~\ref{fig:distributed-cx}.
We note that while the distillation protocol itself requires only one-way communication for error correction, distributed entangling operations, such as the one illustrated in Fig.~\ref{fig:distributed-cx}, may practically necessitate two-way communication, for example, to implement feedforward Pauli corrections.

A challenge that future distillation protocols may face is the rate of Bell pair generation.
Since the codes required for error correction often demand numerous Bell pairs for encoding, one might naively wait for all Bell pairs to be available before starting the encoding process.
This idling time can introduce additional errors on the Bell pairs.
Convolutional codes, such as the SC code, offer a potential solution by allowing encoding to proceed in smaller segments, each of which is itself a code.
This structure enables on-the-fly encoding, which may help mitigate these idling errors.

Focusing on distributed computation, we note that our protocol is compatible with the recent paradigm of algorithmic fault tolerance (AFT) for fast quantum computation~\cite{cain2024correlated,zhou2024algorithmic}.
By leveraging AFT and the transversal operations available in qLDPC codes~\cite{xu2024fast}, we can achieve a factor of $O(d)$ reduction in time cost per transversal logical operation compared to schemes based on lattice surgery~\cite{ramette2023fault,sinclair2024faulttolerant}, which is relevant for the execution of resource-efficient distributed algorithms.

With near-term Bell pair generation rates and infidelities expected to reach values of $10^5 \text{s}^{-1}$ and $p_{\text{bell}} = 0.1\%$, respectively~\cite{li2024high}, qLDPC-based Bell-pair distillation presents a promising approach for high-fidelity quantum interconnects.

Exploring hybrid strategies that first reduce physical Bell-pair error rates via standard distillation before qLDPC encoding, as well as schemes that interpolate flexibly between error detection and error correction using decoder confidence metrics, are compelling directions for further research that may enhance practical performance.

Finally, we emphasize that our distillation protocol is broadly applicable, being compatible with any quantum computing architecture capable of implementing the long-range connectivity required by high-rate qLDPC codes~\cite{ye2025quantum,berthusen2025toward,pattison2025hierarchical,bravyi2024high,yoder2025tour,wang2025demonstration}.

\textit{Acknowledgments.---} We acknowledge helpful discussion with Andi Gu, Senrui Chen, Christopher Pattison, Siyi Yang, Dolev Bluvstein, Nishad Maskara, Marcello Laurel, Rohan Mehta, Patrick Rall, Josiah Sinclair, Nazli Ugur Koyluoglu, Varun Menon.
We acknowledge support from the ARO (W911NF-23-1-0077), ARO MURI (W911NF-21-1-0325, W911NF-20-1-0082), AFOSR MURI (FA9550-19-1-0399, FA9550-21-1-0209, FA9550-23-1-0338), DARPA (HR0011-24-9-0359, HR0011-24-9-0361), NSF (OMA-1936118, ERC-1941583, OMA-2137642, OSI-2326767, CCF-2312755, PHY-2012023, CCF-2313084), NTT Research, Packard Foundation (2020-71479), the Marshall and Arlene Bennett Family Research Program, IARPA and the ARO, under the Entangled Logical Qubits program (W911NF-23-2-0219), the Center for Ultracold Atoms (a NSF Physics Frontiers Center, PHY-1734011), and the NSF Center for Quantum networks (EEC-1941583).
This material is based upon work supported by the U.S. Department of Energy, Office of Science, National Quantum Information Science Research Centers and Advanced Scientific Computing Research (ASCR) program under contract number DE-AC02-06CH11357 as part of the InterQnet quantum networking project.
Q.X. is funded in part by the Walter Burke Institute for Theoretical Physics at Caltech.
G.B. acknowledges support from the MIT Patrons of Physics Fellows Society.

\bibliographystyle{apsrev4-2}
\bibliography{ref}

\clearpage
\newpage

\onecolumngrid
\begin{center}\Large{\textbf{End Matter}}\end{center}
\normalsize
\twocolumngrid

\textit{qLDPC encoder.---} To demonstrate that the ancilla-assisted procedure described in the main text correctly encodes onto the qLDPC code, let the stabilizers of the qLDPC code be given by $\mathcal{S}_{\text{qLDPC}} = \langle g_x \sqcup g_z \rangle$ where  $g_x = \{ g_{x,1}, \dots, g_{x,m}\}$ and $g_z = \{ g_{z,1}, \dots, g_{z,m} \}$.
By performing the encoding procedure, the stabilizers of the new code become:

\vspace{-8mm}
\begin{align*}
    \mathcal{S}' &= \langle g_x' \sqcup g_z' \rangle \\
    g_x' &= \{ g^{\text{A}}_{x,1}\otimes g^{\text{B}}_{x,1}, \dots g^{\text{A}}_{x,m}\otimes g^{\text{B}}_{x,m} \}\\
    g_z' &= \{ g^{\text{A}}_{z,1}\otimes g^{\text{B}}_{z,1} \dots g^{\text{A}}_{z,m}\otimes g^{\text{B}}_{z,m} \}
\end{align*}
 
This holds because, for example, when Alice sends Bob the parity of her first $X$ stabilizer, $g^{\text{A}}_{x,1}$, Bob combines it with the parity of his first $X$ stabilizer, $g^{\text{B}}_{x,1}$, to obtain the value of $g^{\text{A}}_{x,1} \otimes g^{\text{B}}_{x,1}$.
After performing error correction on his side, he projects the joint code state into the $+1$ eigenspace of $g^{\text{A}}_{x,1} \otimes g^{\text{B}}_{x,1}$.
The same logic applies to the remaining stabilizers.
As a result, the stabilizers of the joint code become $\mathcal{S}' = \{ s \otimes s \hspace{1mm} | \hspace{1mm} s \in \mathcal{S}_{\text{qLDPC}}\}$, which is isomorphic to the stabilizers of the qLDPC code: $\mathcal{S}_{\text{qLDPC}} \cong \mathcal{S}'$ via $s \mapsto s\otimes s$.

The logical operators of the code are elements in the centralizer of the stabilizer group.
Since the structure of the qLDPC code’s stabilizers is preserved after the joint projection, the logical operators are also preserved.
Explicitly, suppose the logical operators of the qLDPC code are given by $\mathcal{L}_x = \{\bar{X}_1, \dots \bar{X}_k\}$ and $\mathcal{L}_z = \{\bar{Z}_1, \dots \bar{Z}_k\}$, where $\left[ \bar{X}_i, \bar{Z_j} \right] = 0$ for $i \neq j$ and $\{ \bar{X}_i, \bar{Z_i} \} = 0$.
Due to the isomorphism, the logical operators of the output state are given by

\vspace{-6.5mm}

$$\mathcal{L}'_x = \{ \bar{X} \otimes \bar{X} \hspace{1mm} | \hspace{1mm} \bar{X} \in \mathcal{L}_x\} \hspace{1.5mm} \text{and} \hspace{1.5mm} \mathcal{L}'_z = \{ \bar{Z} \otimes \bar{Z} \hspace{1mm} | \hspace{1mm} \bar{Z} \in \mathcal{L}_z\},$$which are the logical stabilizers of the $k$ output Bell pairs.
We highlight that the encoding protocol is fault-tolerant because the logical Bell stabilizers consist of physical Bell stabilizers that commute with the checks of the joint code, allowing for deterministic fault-tolerant error correction~\cite{wilde2010convolutional}.

\begin{figure*}[h]
    \centering
    \includegraphics[width=\textwidth]{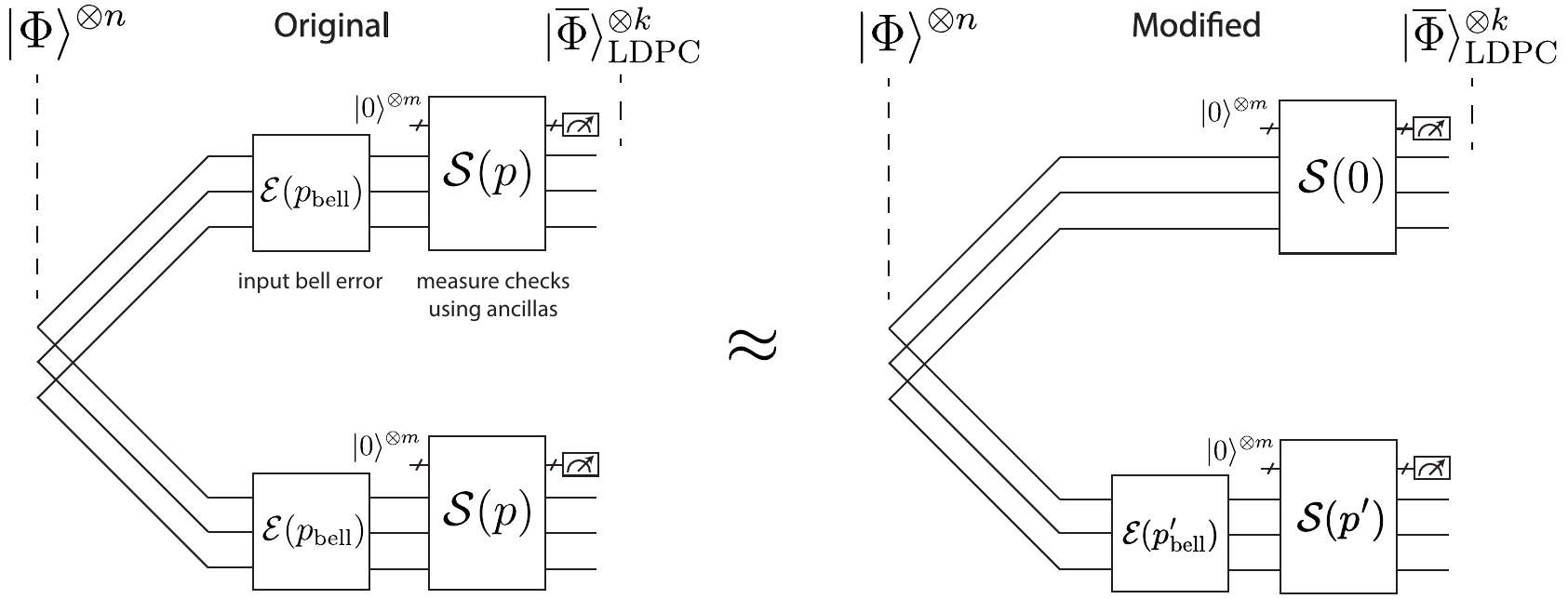}
    \caption{Local and network noise in the encoding protocol.
    Alice and Bob initially share $n$ noisy Bell pairs, modeled by a depolarizing channel with error parameter $p_{\text{bell}}$.
    Syndrome extraction is performed using local ancilla qubits, with a two-qubit gate error rate $p$ at each node.
    To approximate the full noisy circuit (left), where both sides experience noise, we use a simplified model (right) in which Alice’s side is noiseless and Bob’s side has increased noise, with parameters $p'_{\text{bell}} \approx 2p_{\text{bell}}$ and $p' \approx 2p$.
    }
    \label{fig:circuit}
\end{figure*}

\textit{Error Modeling.---}  
The error channels are illustrated in Fig.~\ref{fig:circuit} (left).  
Each Bell pair undergoes independent depolarizing noise at each node, modeled as 
\begin{equation}
    \mathcal{E}(p_{\text{bell}}) = (1 - p_{\text{bell}}) \rho + \frac{p_{\text{bell}}}{3} (X\rho X + Y\rho Y + Z\rho Z).
\end{equation}  

The QEC circuits experience two-qubit gate depolarizing errors with strength \( p \), given by  

\begin{equation}
    \mathcal{S} (p) = (1 - p) \rho + \frac{p}{15} \sum_{P \in \{IX, IY, \dots, ZZ\}} P \rho P.
\end{equation}  

To simplify simulations, we shift errors from Alice's side to Bob's, effectively increasing the noise on Bob’s side.
This allows us to simulate only Bob’s end, where the effective Bell error rate and circuit error rate are denoted \( p'_{\text{bell}} \) and \( p' \) (Fig.~\ref{fig:circuit}, right).
In the Supplementary Information~\cite{SM}, we use the symmetry of Bell pairs under Pauli operators to show that: $p_{\text{bell}}' \approx 2p_{\text{bell}}$.
As discussed in the main text, we quote $p_{\text{bell}}'$ as the input Bell error rate.

Similarly, we relate the two-qubit gate error rates of the QEC circuit, \( p \) and \( p' \), by considering physical error events that lead to logical errors on the encoded Bell pairs.
In the Supplementary Information, we show that $p' \approx 2 p$ is a valid approximation for single-round QEC circuits.
We also numerically verify that, for a single round of QEC, joint decoding yields approximately the same threshold as decoding only Bob's side using the effective error model.
For a large number of QEC rounds, \( p' \approx 2p \) becomes a conservative, as argued in the SI.
Nonetheless, we use \( p' = 2p \) for the simulations in the main text, where the local gate error is fixed at \( 0.1\% \), resulting in an effective simulated gate error of \( p' = 0.2\% \).

\textit{Codes.---}
The HGP codes are constructed as the hypergraph product of two classical LDPC codes based on (3,4)-regular Tanner graphs~\cite{xu2024constant,tremblay2022constant,grospellier2021combining}.
The LP codes extend this construction by incorporating a lifting operation, which reduces qubit overhead while preserving high code distances~\cite{panteleev2019degenerate,breuckmann2020balanced,raveendran2022finite}.
Finally, the SC codes build upon the LP structure by introducing spatial coupling of the lifted component matrices -- a form of parity-check matrix concatenation -- which further enhances qubit efficiency, achieving an encoding rate of $\sim 1/3$~\cite{hagiwara2011spatially,yang2023spatially}. 
All codes are optimized to (1)~reduce small loops in the Tanner graph, which improves Belief Propagation decoding, (2)~enhance expansion properties, facilitating single-shot decoding, and (3)~maximize distance, improving error correction performance.
Further details on code constructions are provided in the Supplementary Information~\cite{SM}.

\textit{Resource requirements.---}
To investigate the sub-threshold scaling behavior of the codes, we modify the error model so that both the Bell-pair error and circuit error scale with the noise strength.
We set $p_{\text{bell}} = p, \quad p_{\text{2q}} = \frac{p}{50}$ so that the observed thresholds in the main text remain approximately preserved.
We then vary \( p \) and fit the data to a sub-threshold scaling ansatz.
The ansatz we use is $\bar{P} = A \left(\frac{p}{p_{\text{th}}}\right)^{Bn^C}$~\cite{xu2024constant}, where \( A, B, C \) are fitting parameters, \( n \) is the code size, and \( p_{\text{th}} \) is the threshold physical error rate (also fitted).
The numerical data collected and the fitted ansatz are shown in Fig.~\ref{fig:subthr}.
The fitted logical error rates are given by: $ \bar{P}_{\text{HGP}} = 0.07 \left(\frac{p}{0.11}\right)^{0.31n^{0.38}}, \bar{P}_{\text{LP}} = 0.026 \left(\frac{p}{0.094}\right)^{0.05n^{0.79}}, \bar{P}_{\text{SC}} = 0.26 \left(\frac{p}{0.087}\right)^{0.26n^{0.44}}.
$
We document achievable logical Bell error rates as a function of input Bell error rates for select codes in Table~\ref{tab:ler}.

\sisetup{exponent-product=\times, output-exponent-marker=\ensuremath{\mathrm{e}}, 
         tight-spacing=true, detect-weight=true, detect-inline-weight=math} 

\begin{table*}[h]
    \centering
    \resizebox{\textwidth}{!}{
    \begin{tabular}{c S[scientific-notation=true] S[scientific-notation=true] S[scientific-notation=true] S[scientific-notation=true] S[scientific-notation=true] S[scientific-notation=true] S[scientific-notation=true] S[scientific-notation=true] S[scientific-notation=true] S[scientific-notation=true]}
    \toprule
    & \multicolumn{10}{c}{Input Bell error} \\
    \cmidrule(lr){2-11}
    \makecell{Code \\ (Overhead)} & \textbf{1\%} & \textbf{2\%} & \textbf{3\%} & \textbf{4\%} & \textbf{5\%} & \textbf{6\%} & \textbf{7\%} & \textbf{8\%} & \textbf{9\%} & \textbf{10\%} \\
    \midrule
    \makecell{$[[2500,100]]$ \\ ($O=25$)} & \num{6.0e-9}* & \num{7.6e-6} & \num{4.2e-5} & \num{2.0e-4}  & \num{6.9e-4} & \num{1.9e-3}  & \num{4.4e-3} & \num{9.3e-3}  & \num{1.9e-2} & \num{3.6e-2}  \\
    \midrule
    \makecell{$[[1428,184]]$ \\ ($O=7.8$)} & \num{2.0e-17}* & \num{9.5e-13}* & \num{5.1e-10}* & \num{4.5e-8}* & \num{4.0e-6} & \num{2.8e-5} & \num{2.8e-4} & \num{2.5e-3} & \num{1.1e-2} & \num{3.2e-2} \\
    \midrule
    \makecell{$[[5800,1624]]$ \\ ($O = 3.6$)} & \num{2.3e-12}* & \num{7.9e-9}* & \num{9.4e-7}* & \num{2.8e-5} & \num{1.9e-4}  & \num{1.9e-3} & \num{1.7e-2} &  \num{8.7e-2} &  &  \\
    \bottomrule
    \end{tabular}
    }
    \caption{Error rates of encoded Bell pairs as a function of input Bell error rates.
    Values marked with * represent extrapolated error rates obtained by fitting to subthreshold data.
    Unmarked values are obtained from exact numerical simulations.
    The overhead (\( O \)) indicates the number of physical Bell pairs required per logical Bell pair.
    }
    \label{tab:ler}
\end{table*}

\renewcommand{\thefigure}{S\arabic{figure}}
\setcounter{secnumdepth}{3}
\renewcommand\thesection{\arabic{section}}

\newpage
\clearpage
\onecolumngrid

\begin{center}\Large{\textbf{Supplementary Materials}}\end{center}
\normalsize

\tableofcontents

\newpage

\section{Related work}

We expand on several relevant Bell-pair distillation methods using the criteria introduced in the main text.

\begin{itemize}
    \item \textbf{BDSW-1EPP}~\cite{bennett1996mixed}: A one-way ``hashing" method capable of achieving a constant rate equal to the Hashing bound of the channel.
    The rate is given by $R = 1 - H(p)$, where the Von-Neumann entropy of the error channel is: $$H\left(p\right)=-\left(1-p\right)\log_{2}\left(1-p\right)-p\log_{2}\left(\frac{p}{3}\right).$$
    At a Bell-Pair infidelity of $5\%$ ($1\%$), the rate is 0.63 (0.90).
    The protocol relies on random quantum codes, for which no efficient decoding methods are currently known, rendering it impractical with existing tools.
    \item \textbf{BDSW-2EPP}~\cite{bennett1996mixed}: A two-way distillation method with a vanishing asymptotic rate.
    The protocol can be considered as a subset of Ref.~\cite{pattison2024fast}, where the concatenated code is the [2,1,2] classical repetition code in alternating $X$ and $Z$ bases.
    The scheme is not robust against gate errors, as it un-encodes to physical Bell pairs.
    \item \textbf{Ramette et al.}~\cite{ramette2023fault}: A lattice-surgery-based one-way distillation scheme.
    Since the scheme is based on surface codes it has a vanishing rate.
    At an input Bell infidelity of $5\%$ ($1\%$), the rate is lower than 1/5300 (1/1300) to achieve a target logical error rate of $10^{-12}$~\cite{pattison2024fast}.
    A total of $O(d^2)$ Bell pairs must be shared at the boundary for surface codes of distance $d$.
    The lattice surgery protocol requires a circuit of depth $d$, where each step consumes $d$ Bell pairs.
    The boundary threshold, where the Bell pairs reside, is approximately $10\%$, while the bulk threshold, where local qubits reside, is approximately $1\%$.
    Other lattice surgery proposals~\cite{fowler2010surface,sinclair2024faulttolerant,leone2024upper} share similar performance metrics.
    \item \textbf{Shi et al.}~\cite{shi2024stabilizer}: A fault-tolerant encoder with a non-fault-tolerant un-encoder for qLDPC codes.
    This scheme requires two-way communication: Alice sends data to Bob for encoding, and Bob sends data back to Alice for decoding.
    The method does not require additional qubits beyond those of the qLDPC code (and potentially some local ancillas).
    However, it does not consider specific high-rate qLDPC codes, and no threshold simulations are performed.
    \item \textbf{Pattison et al.}~\cite{pattison2024fast}: A two-way error-detecting method that achieves a constant rate through concatenation of small codes.
    Due to concatenation and post-selection, this method may require a large buffer memory, as faulty Bell pairs are discarded.
    The resulting encoding memory overhead scales as \(O\left( (\log \log 1/\epsilon)^{\alpha \log \log 1/\epsilon} \right) > O(1),\)
    for an output error rate \( \epsilon \) and some \( \alpha > 0 \).
    At an input Bell infidelity of $5\%$ ($1\%$), and assuming a buffer memory of 50 logical qubits, the protocol achieves a rate of 16.53 (7.32) to reach a target logical error rate of $10^{-12}$.
    \item \textbf{Gu et al.}~\cite{gu2025constant} (concurrent with our work): A two-way error-detecting method that achieves a constant rate using random bilocal Clifford circuits.
    The scheme requires a buffer memory of \(\log_2(1/\epsilon)^{3/2} \text{ qubits}\) for each Alice and Bob, for an output error rate \( \epsilon \).
    Assuming noiseless local operations, it can achieve an overhead of 7 at an input Bell-pair infidelity of 10\%, reaching a target logical error rate of $10^{-12}$.
    The scheme can be made robust to local gate errors by injecting Bell pairs into a QEC code at each node.
    \item \textbf{This work}: A deterministic (one-way) constant-rate distillation protocol based on high-rate qLDPC codes.
    The resource overhead is constant, as the scheme requires no additional Bell pairs beyond those used in the qLDPC block, along with \( O(1) \) local ancillas.
    The output Bell pairs remain encoded in the qLDPC code, making the scheme robust against local gate noise.
    Since the same qLDPC code is used for both distillation and local gate noise suppression, the entire fault-tolerant pipeline can be simulated.
    We analyze various schemes:
    \begin{itemize}
        \item A family of HGP codes with a rate \( \geq 4\% \) and a threshold of \( 10\% \), below which the Bell-pair error can be arbitrarily suppressed.
        \item A family of LP codes with a rate \( \sim 11\% \), with four finite-size instances constructed, and a threshold of \( \sim 10\% \).
        \item A rate \( \sim 1/3 \) code with a pseudothreshold of \( 7.5\% \), achieving a logical error rate of \( \sim 10^{-12} \) at an input Bell error rate of \( 1\% \) (extrapolated from the subthreshold fitted ansatz).
    \end{itemize}
    See the End Matter of the main text for full results.
\end{itemize}

We note that while various other purification methods exist~\cite{bennett1996purification,murao1998multiparticle,vollbrecht2005interpolation,isailovic2006interconnection,hostens2006asymptotic,leung2007entanglement,hayden2008decoupling,leung2008quantum,krastanov2019optimized,gidney2023tetrationally,goodenough2024bipartite,rengaswamy2024entanglement,siddhu2024entanglement,shi2025measurementbased}, a detailed comparison of these methods is left for future work.

\section{Noise manipulation}

In this section, we analyze how Bell-pair depolarizing noise and circuit noise transform under different setups, leading to effective noise models for our numerical simulations.

\subsection{Depolarizing noise}

Consider a Bell pair with one qubit at Alice's node and the other at Bob's node:
$$\ket{\Phi^+} = \frac{1}{\sqrt{2}} (\ket{0}_A\otimes \ket{0}_B + \ket{1}_A \otimes \ket{1}_B)$$

Each qubit undergoes a depolarizing channel with error probability $p_{\text{bell}}$:

$$\mathcal{E}^{\text{A}}_{p}(\rho_{\text{AB}}) = (1-p_{\text{bell}})\rho_{\text{AB}} + \frac{p_{\text{bell}}}{3} \sum_{P\in \{X,Y,Z\}} P_{\text{A}}\rho_{\text{AB}}P_{\text{A}},$$

$$\mathcal{E}^{\text{B}}_{p}(\rho_{\text{AB}}) = (1-p_{\text{bell}})\rho_{\text{AB}} + \frac{p_{\text{bell}}}{3}\sum_{P\in \{X,Y,Z\}} P_{\text{B}}\rho_{\text{AB}}P_{\text{B}}.$$

\begin{figure*}[h]
    \centering
    \includegraphics[width=0.7\textwidth]{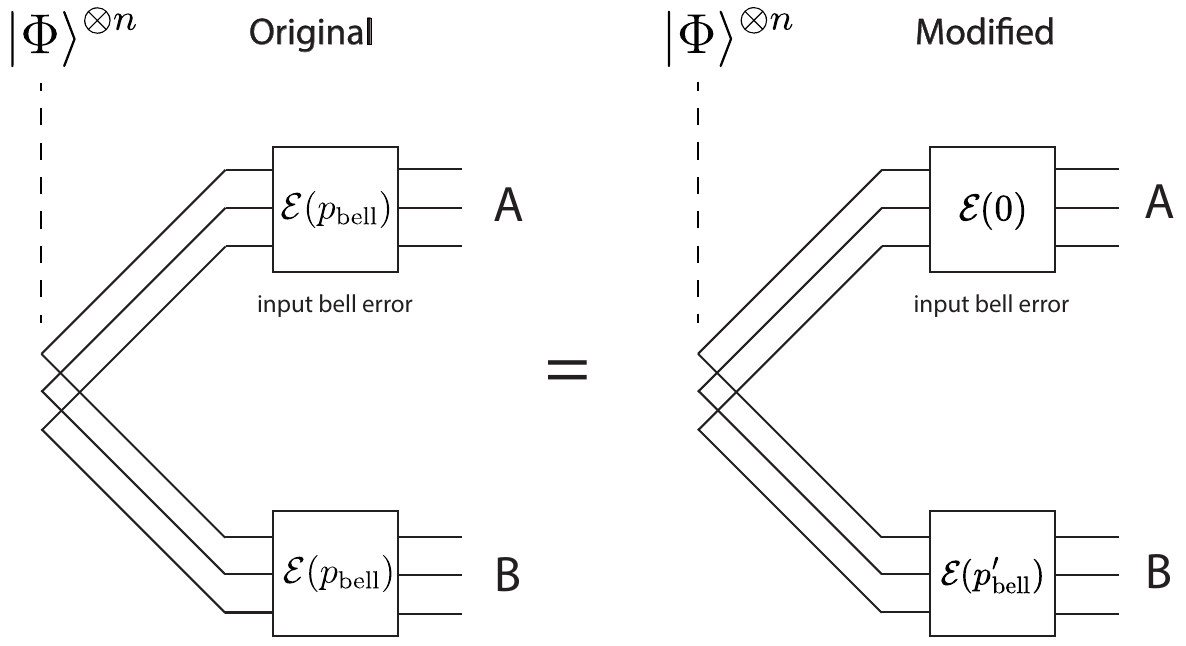}
    \caption{Bell error equivalence.
    The left setup represents the original configuration where both Alice and Bob experience input Bell pair depolarizing noise $\mathcal{E}(p_{\text{bell}})$.
    The right setup illustrates the equivalent modified configuration where Alice’s side is noiseless $\mathcal{E}(0)$, and Bob’s side has an effective increased depolarizing error $\mathcal{E}(p_{\text{bell}}')$.
    }
    \label{fig:bell-noise}
\end{figure*}

We seek an equivalent noise model where Alice’s side is noiseless, and Bob’s side has an effective depolarizing error $p_{\text{bell}}'$ (Fig.~\ref{fig:bell-noise}).
In Table~\ref{tab:bell-noise}, we enumerate all possible Pauli operator configurations resulting from the depolarizing channels on Alice's and Bob's sides and the corresponding resulting Bell states.
By summing the probabilities of these configurations, we determine the likelihood of each Bell state for both the original (left) and modified (right) setups.
Equating the probabilities of each Bell state in the two setups provides a consistent solution for the effective error parameter: $p_{\text{bell}}' = 2p_{\text{bell}} - \frac{4}{3}p_{\text{bell}}^2$.

\begin{table}[h]
\centering
\resizebox{\textwidth}{!}{
\begin{tabular}{|c|c|c|c|c|}
\hline
Bell state & Error Mechanisms (original) & Probability (original) & Error Mechanisms (modified) & Probability (modified) \\ \hline
$\ket{\Phi^+}$ & $I_{\text{A}}I_{\text{B}}, X_{\text{A}}X_{\text{B}}, Y_{\text{A}}Y_{\text{B}}, Z_{\text{A}}Z_{\text{B}}$ & $(1-p_{\text{bell}})^2 + \frac{p_{\text{bell}}^2}{3}$ & $I_{\text{B}}$ & $1-p_{\text{bell}}'$ \\ \hline
$\ket{\Phi^-}$ & $I_{\text{A}}Z_{\text{B}}, Z_{\text{A}}I_{\text{B}}, X_{\text{A}}Y_{\text{B}}, Y_{\text{A}}X_{\text{B}}$ & $\frac{2p_{\text{bell}}}{3}(1-p_{\text{bell}}) + \frac{2p_{\text{bell}}^2}{9}$ & $Z_{\text{B}}$ & $\frac{p_{\text{bell}}'}{3}$ \\ \hline
$\ket{\Psi^+}$ & $I_{\text{A}}X_{\text{B}}, X_{\text{A}}I_{\text{B}}, Y_{\text{A}}Z_{\text{B}}, Z_{\text{A}}Y_{\text{B}}$ & $\frac{2p_{\text{bell}}}{3}(1-p_{\text{bell}}) + \frac{2p_{\text{bell}}^2}{9}$ & $X_{\text{B}}$ & $\frac{p_{\text{bell}}'}{3}$ \\ \hline
$\ket{\Psi^-}$ & $I_{\text{A}}Y_{\text{B}}, Y_{\text{A}}I_{\text{B}}, X_{\text{A}}Z_{\text{B}}, Z_{\text{A}}X_{\text{B}}$ & $\frac{2p_{\text{bell}}}{3}(1-p_{\text{bell}}) + \frac{2p_{\text{bell}}^2}{9}$ & $Y_{\text{B}}$ & $\frac{p_{\text{bell}}'}{3}$ \\ \hline
\end{tabular}
}
\caption{Pauli operators and the resulting Bell states.
Summary of the Pauli operators acting on Alice's and Bob's sides, their corresponding Bell states, and the probabilities of each configuration under the depolarizing noise model for the original and modified setups.
The probabilities of the Bell states in the original and modified setups are equated to derive the effective error parameter $p_{\text{bell}}' = 2p_{\text{bell}} - \frac{4}{3}p_{\text{bell}}^2$.}
\label{tab:bell-noise}
\end{table}

\subsection{Circuit noise}

\begin{figure*}[h]
    \centering
    \includegraphics[width=0.6\textwidth]{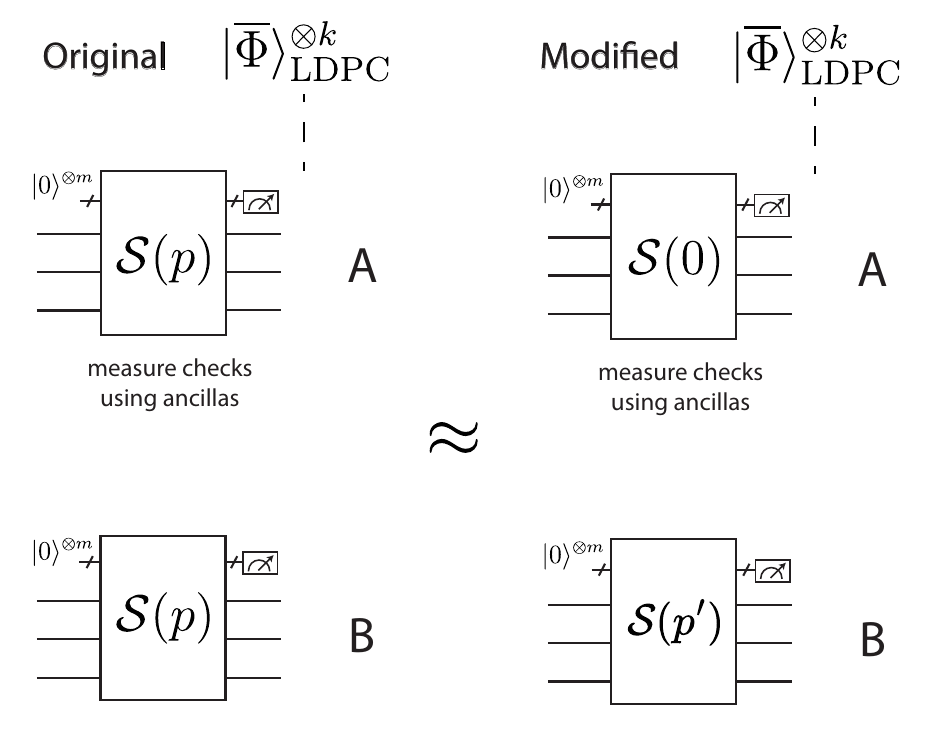}
    \caption{Circuit error approximation.
    The left setup shows the original configuration, where both Alice and Bob's circuits experience errors characterized by $\mathcal{S}$.
    The right setup illustrates the modified configuration, where Alice’s circuit is noiseless $\mathcal{S}(0)$, and Bob’s circuit has an effective increased error strength $\mathcal{S}(p')$.
    }
    \label{fig:circuit-noise}
\end{figure*}

Suppose we run the same circuit on each leg of the Bell pair.
The error channel is parametrized by an error strength $p$, such that physical operations in the circuit fail with probability $p$, or some constant factor of $p$.
Consider the two setups shown in Fig.~\ref{fig:circuit-noise}.
On the left, the error channels on Alice's and Bob's sides are identical, each with the error parameter $p$.
On the right, Alice's circuit is noiseless, and Bob's circuit has an effective error parameter $p'$.
Our goal is to relate $p'$ to $p$, using reasoning similar to the previous section.

The circuits are error correction circuits, where syndrome extraction is performed using bare, local ancilla qubits, as described in the main text.
The logical operators and resulting logical Bell states are consistent with those presented in Table~\ref{tab:bell-noise}.

For this analysis, we focus on logical operators that result in a $\ket{\bar{\Phi}^-}$ state, though the results apply to any row in Table~\ref{tab:bell-noise}.
Specifically, we seek to relate $p'$ and $p$ under the condition that a logical $\bar{Z}$ error occurs.
In the modified setup, the probability of a logical error, $\mathbb{P}_p(\bar{Z})$, at sufficiently low physical error rates is given by summing over all possible error configurations.
This can be expressed as a sum over circuit fault weights $\omega$, with an entropy factor, $A_{\omega}$ accounting for the number of weight-$\omega$ error configurations that lead to a logical failure:

$$\mathbb{P}_{p'}(\bar{Z}) \approx \sum_{\omega} A_{\omega} (p')^{\omega}.$$

In the original setup, each weight-$\omega$ error configuration that leads to failure in the modified setup also leads to failure.
However, there is an additional degeneracy: each physical error can occur on either Alice’s or Bob’s side, yet the logical effect remains the same due to the joint decoding of Alice's and Bob's syndromes.
For example, consider a 4-qubit repetition code, where $ZZII$ results in a logical $\bar{Z}$ error.
In the joint code (with stabilizers $s\otimes s$, where $s$ is a stabilizer in the original code, as derived in the main text), the modified setup allows only the configuration: $$(IIII) \otimes (ZZII).$$
In contrast, the original setup exhibits a $2^2$-fold degeneracy as the following configurations all lead to a logical $\bar{Z}$ error (ignoring higher weight corrections~\footnote{We note that there are higher weight terms that can also lead to failure - for example, $(XZII)\otimes(YIII)$.
We ignore these higher weight corrections when deriving our approximation for $p'$.}): 

$$(IIII)\otimes(ZZII) \quad (ZZII)\otimes(IIII) \quad (ZIII)\otimes(IZII) \quad (IZII)\otimes(ZIII).$$

More generally, for any weight-$\omega$ error configuration in the modified setup, there are $2^\omega$ equivalent weight-$\omega$ configurations in the original setup.
Thus, the probability of a logical $\bar{Z}$ error in the original setup, ignoring higher weight corrections, is:

$$\mathbb{P}_{p}(\bar{Z}) \approx \sum_{\omega} A_{\omega} 2^\omega p^{\omega}.$$

By equating the probabilities $\mathbb{P}_{p'}(\bar{Z}) = \mathbb{P}_{p}(\bar{Z})$ we obtain: $$p' \approx 2p.$$

At very low physical error rates, where weight-$\sim d/2$ error configurations dominate, our approximations becomes increasingly tight.

\subsection{Generalizations}

\subsubsection{Combination of Bell and Circuit Errors}

To incorporate both Bell depolarizing errors and circuit gate errors into a unified framework, we analyze the entire protocol within the circuit error model.
In this framework, the Bell depolarizing error acts as a preparation error on the input qubits, which can equivalently be represented as perfect initialization followed by a Pauli error.
Thus, the results from the previous sections apply simultaneously to both Bell depolarizing and gate errors, justifying the use of $p'\approx 2p$ and $p_{\text{bell}}' \approx 2p_{\text{bell}}$.

\subsubsection{Generalization to Multiple Cycles}

In the limit of a large number of QEC cycles on both sides, the system effectively behaves as two independent logical qubits.
For such independent logical qubits, the logical error rate per round scales as: $$f(p') = 2\times f(p),$$ whereas our approximation assumes $$f(p') = f(2p),$$ which provides a conservative upper bound on the actual error rate.
This holds because: $$f(p) = (p/p_{\text{th}})^{d/2}$$ is a high power function.
Therefore, $p' \approx 2p$ provides a reasonable conservative estimate for the case of multiple QEC cycles.

\section{Code constructions}

\subsection{Hypergraph product code}

HGP codes are a class of quantum low-density parity-check (qLDPC) codes derived from the Cartesian product of two classical LDPC codes~\cite{tillich2014quantum}.
Let the parity-check matrices of the classical codes be \( H_1 \in \mathbb{F}_2^{r_1 \times n_1} \) and \( H_2 \in \mathbb{F}_2^{r_2 \times n_2} \), where \( r_1, r_2 \) are the number of checks, and \( n_1, n_2 \) are the number of bits.
The quantum stabilizer matrices are given by:
\[
H_X = \begin{bmatrix}
H_1^T \otimes I_{r_2} & I_{n_1} \otimes H_2
\end{bmatrix}, \quad
H_Z = \begin{bmatrix}
I_{r_1} \otimes H_2^T & H_1 \otimes I_{n_2}
\end{bmatrix},
\]
where \( \otimes \) denotes the Kronecker product.
This construction guarantees the CSS condition \( H_X H_Z^T = 0 \).

For classical codes with parameters \([n_i, k_i, d_i]\), where $r_i = n_i - k_i$ are the linearly-independent checks, the resulting HGP code encodes \( k = k_1 k_2 \) logical qubits into \( n = n_1 n_2 + r_1 r_2 \) physical qubits, with distance \( d = \min(d_1, d_2) \).
If the classical codes are repetition codes \([n,1,O(n)]\), the HGP construction yields the surface code.
If, instead, the classical codes encode a constant number of logical bits \([n,O(n),O(n)]\), the HGP construction yields constant-rate codes with square-root distance \( [[n, O(n), O(\sqrt{n})]] \).

In this work, the classical codes are constructed using (3,4)-regular Tanner graphs~\cite{xu2024constant,tremblay2022constant,grospellier2021combining}, which are bipartite graphs where each bit node has degree 3 and each check node has degree 4.
We generate candidate (3,4)-regular Tanner graphs via rejection sampling, selecting the best one based on:
\begin{itemize}
    \item \textbf{High girth} (\(\geq 6\)) to minimize short cycles, improving decoder performance.
    \item \textbf{Large spectral gap}, which enhances expansion properties and is associated with single-shot QEC.
    \item \textbf{Maximum code distance}, ensuring improved error correction performance.
\end{itemize}

By varying the size of the Tanner graph, we construct a family of classical codes \( C_1, C_2, \dots \), from which we derive a family of quantum codes by taking the hypergraph product of the classical code with itself, $Q=$HGP$(C,C)$:

\[
Q_1 = [[225,9,4]],\quad Q_2=[[625,25,6]], \quad Q_3 = [[1225,49,8]],\quad  \dots
\]

These codes achieve a minimum rate of \( k/n \geq 4\% \).
The HGP codes used in this article are the same as those studied in Ref.~\cite{xu2024constant}.

The structure of HGP codes lends itself to a compact implementation protocol using reconfigurable atom arrays~\cite{xu2024constant}.
Since lifted-product (LP) and spatially-coupled (SC) codes are built using HGP codes, the optimized movement schemes developed for HGP codes can also aid the implementation of the LP and SC codes.

\subsection{Lifted product code}

Quasi-cyclic lifted Product (LP) codes are a family of quantum low-density parity-check (qLDPC) codes that enhance the hypergraph product by introducing a lifting operation~\cite{panteleev2019degenerate,breuckmann2020balanced,raveendran2022finite}.
This operation uses cyclic group structures to reduce the number of required qubits while maintaining high code rates and robust error-correcting properties.

LP codes are constructed using two classical base protographs, represented by matrices \( B_1 \) and \( B_2 \) over the quotient polynomial ring \( R[x]/(x^l - 1) \).
These matrices generate larger matrices \( B_x \) and \( B_z \):

\[
B_x = \begin{bmatrix}
B_1^T \otimes I_{m_{B2}} & I_{n_{B1}} \otimes B_2
\end{bmatrix}, \quad
B_z = \begin{bmatrix}
I_{m_{B1}} \otimes B_2^T & B_1 \otimes I_{n_{B2}}
\end{bmatrix}.
\]

Here, \( l \) denotes the lift size, and \( B_1 \), \( B_2 \) correspond to the underlying protographs of size \( m_{B1} \times n_{B1} \) and \( m_{B2} \times n_{B2} \), respectively.
The ``lifting" operation replaces each element of \( B_x \) and \( B_z \) with its corresponding circulant \( l \times l \) matrix representation.
This process generates the \( H_X \) and \( H_Z \) stabilizer matrices required for the quantum code, ensuring that \( H_X H_Z^T = 0 \).

The LP code has parameters \( [[n, k, d]] \), where:
\begin{itemize}
    \item \( n = l(n_{B1} n_{B2} + m_{B1} m_{B2}) \),
    \item \( k  = l(n_{B1} n_{B2} + m_{B1} m_{B2} - m_{B1} n_{B2} - n_{B1} m_{B2}) \), and
    \item \( d \) is the minimum distance, upper bounded by the classical distance of the lifted base matrix.
\end{itemize}

We construct LP codes using base matrices \( B_1 \) and \( B_2 \) with monomial entries and size \( 3 \times 5 \).
To optimize the LP codes, the base matrices are selected to maximize girth (\(\geq 8\)) and distance.
To form a code family, we vary the lift size \( l = 16, 21, 30, 42 \), obtaining base matrices \( B_1, B_2, B_3, B_4 \) with classical distances \( d = 12, 16, 20, 24 \):

\[
\mathbf{B}_1 =
\begin{bmatrix}
1 & 1 & 1 & 1 & 1 \\
1 & x^2 & x^4 & x^7 & x^{11} \\
1 & x^3 & x^{10} & x^{14} & x^{15}
\end{bmatrix}, \quad
\mathbf{B}_2 =
\begin{bmatrix}
1 & 1 & 1 & 1 & 1 \\
1 & x^4 & x^5 & x^7 & x^{17} \\
1 & x^{14} & x^{18} & x^{12} & x^{11}
\end{bmatrix}, \quad
\mathbf{B}_3 =
\begin{bmatrix}
1 & 1 & 1 & 1 & 1 \\
1 & x^2 & x^{14} & x^{24} & x^{25} \\
1 & x^{16} & x^{11} & x^{14} & x^{13}
\end{bmatrix}, \quad
\mathbf{B}_4 =
\begin{bmatrix}
1 & 1 & 1 & 1 & 1 \\
1 & x^6 & x^7 & x^9 & x^{30} \\
1 & x^{40} & x^{15} & x^{31} & x^{35}
\end{bmatrix}
\]

and associated quantum code parameters $[[544,80,\leq12]], \hspace{1mm} [[714,100,\leq 16]], \hspace{1mm} [[1020,136,\leq 20]], \hspace{1mm} [[1428,184,\leq 24]]$. 

The constructed LP codes achieve encoding rates lower-bounded by \( 2/17 \) and distances matching the underlying classical matrices with high probability.
This LP code construction follows the same method used in Ref.~\cite{raveendran2022finite} and Ref.~\cite{xu2024constant}.

\subsection{Spatially-coupled code}

Spatially Coupled Quantum Low-Density Parity-Check (SC-qLDPC) codes extend classical spatially coupled (SC) codes to the quantum setting~\cite{hagiwara2011spatially,yang2023spatially}.
Following Ref.~\cite{yang2023spatially}, we outline the construction of the SC codes used in this work.

At a high level, SC codes can be viewed as LP codes with an additional coupling structure, where the lifted matrices are stacked vertically and horizontally in a repeating pattern.
This procedure reduces excess physical qubits while maintaining a large number of encoded logical qubits, thereby increasing the code rate.
A partitioning matrix \( \mathbf{P} \) is used to decompose the base matrix \( \mathbf{B} \) into \( m+1 \) component matrices, each of which is lifted according to a lifting matrix \( \mathbf{L} \), generating the lifted component matrices \( H_i \) (\( i=0,1,\dots,m \)).
These component matrices are then stacked vertically to form a so-called replica, and the replicas are stacked horizontally to form the full check matrix \( H \).
For a class of SC codes known as tail-biting (TB) codes, the resultant check matrix is given by:

\[
\mathbf{H} =
\begin{bmatrix}
    H_0 & 0 & \cdots & 0 & H_m & \cdots & H_1 \\
    H_1 & H_0 & 0 & \cdots & 0 & \ddots & \vdots \\
    \vdots & H_1 & H_0 & \ddots & \vdots & \ddots & H_m \\
    H_m & \vdots & \ddots & 0 & 0 & \cdots & 0 \\
    0 & H_m & \ddots & H_1 & H_0 & \ddots & \vdots \\
    \vdots & \ddots & \ddots & \vdots & H_1 & \ddots & 0 \\
    0 & \cdots & 0 & H_m & \cdots & H_1 & H_0
\end{bmatrix}
\]

Here, \( m \) is known as the memory of the SC code, and the coupling length \( L \) refers to the number of columns of \( H \).

A Two-Dimensional (2D) SC code extends the 1D SC structure by coupling multiple SC codes together.
A memory-\( m_1 \) SC code is constructed from \( m_1+1 \) component matrices, each of which is itself a memory-\( m_2 \) SC code.
The final code is characterized by outer and inner coupling lengths, \( L_1 \) and \( L_2 \), respectively.
The check matrix is then specified by check matrices \( H_{ij} \), where \( i=0,1,\dots,m_1 \) and \( j=0,1,\dots,m_2 \).

The Toric code is an example of a 2D-SC code, where $(m_1,m_2,L_1,L_2)=(1,1,d,d)$.
For instance, the \( d=3 \) Toric code has parity-check matrix:

\[
\mathbf{H}_{\text{Toric}} =
\begin{bmatrix}
    \begin{array}{c|c|c}
        \begin{array}{ccc}
            A &  & B \\
            B & A &  \\
             & B & A
        \end{array} &
        \begin{array}{ccc}
             & &  \\
             & & \\
             & &
        \end{array} &
        \begin{array}{ccc}
            C & & D \\
            D & C & \\
            & D & C
        \end{array} \\
        \hline
        \begin{array}{ccc}
            C & & D \\
            D & C & \\
            & D & C
        \end{array} &
        \begin{array}{ccc}
            A &  & B \\
            B & A &  \\
             & B & A
        \end{array} &
        \begin{array}{ccc}
             & &  \\
             & & \\
             & &
        \end{array} \\
        \hline
        \begin{array}{ccc}
             & &  \\
             & & \\
             & &
        \end{array} &
        \begin{array}{ccc}
            C & & D \\
            D & C & \\
            & D & C
        \end{array} &
        \begin{array}{ccc}
            A &  & B \\
            B & A &  \\
             & B & A
        \end{array}
    \end{array}
\end{bmatrix},
\]

where the block matrices are:

\[
\mathbf{A} =
\begin{bmatrix}
    X & I \\
    I & I
\end{bmatrix}, \quad
\mathbf{B} =
\begin{bmatrix}
    X & X \\
    Z & I
\end{bmatrix}, \quad
\mathbf{C} =
\begin{bmatrix}
    I & X \\
    Z & Z
\end{bmatrix}, \quad
\text{and} \quad
\mathbf{D} =
\begin{bmatrix}
    I & I \\
    I & Z
\end{bmatrix}.
\]

\subsubsection{Algebraic formulation}

The algebraic formulation for SC codes facilitates the construction of the code we use.
At its core is the characteristic function, \( F(U,V) \), which determines the component matrices \( H_{ij} \) of the final parity-check matrix:

\[
F(U,V) = \sum_{i=0}^{m_1} \sum_{j=0}^{m_2} H_{ij} U^i V^j.
\]

For 2D SC-HGP codes, the characteristic function takes a form similar to the hypergraph product:

\[
\mathbf{F}(U,V) =
\begin{bmatrix}
    X \big( I_{n_2 \times n_2} \otimes \mathbf{A}(U,V) \big) & X \big( \bar{\mathbf{B}}(U,V)^T \otimes I_{r_1 \times r_1} \big) \\
    Z \big( \mathbf{B}(U,V) \otimes I_{n_1 \times n_1} \big) & Z \big( I_{r_2 \times r_2} \otimes \bar{\mathbf{A}}(U,V)^T \big)
\end{bmatrix},
\]

where \(\mathbf{A}(U,V) \in \mathbb{F}_2^{r_1 \times n_1}[U,V]\) and 
\(\mathbf{B}(U,V) \in \mathbb{F}_2^{r_2 \times n_2}[U,V]\).
One may interpret \( \mathbf{F}(U,V) \) as the hypergraph product of \( \mathbf{A} \) and \( \mathbf{B} \).
For each of \( \mathbf{A} \) and \( \mathbf{B} \), there exists an associated partitioning matrix \( \mathbf{P}_a \) and \( \mathbf{P}_b \) that fully determine the final partitioning matrix \( \mathbf{P} \), which itself is the hypergraph product of \( \mathbf{P}_a \) and \( \mathbf{P}_b \).

To construct the final SC-HGP code, we begin with base matrices $\textbf{A}$ and $\textbf{B}$ of size $r_1\times n_1$ and $r_2 \times n_2$, respectively, memories $m_1$ and $m_2$ and coupling lengths $L_1$ and $L_2$.
Applying the hypergraph product, the resulting check matrix has $r=r_1n_2 + r_2n_1$ rows and $n=r_1r_2 + n_1n_2$ columns, encoding at least $k=n-r = (n_1-r_1)(n_2-r_2)$ logical qubits.
After spatial coupling, the final matrix has $N = (r_1r_2 + n_1n_2)L_1L_2$ physical qubits and encodes at least $K=(n_1-r_1)(n_2-r_2)L_1L_2$ logical qubits.

For the SC-HGP code construction used in this paper, the following parameters are used:

\[
r_1 = r_2 = 3, \quad n_1 = n_2 = 7, \quad m_1 = m_2 = 3, \quad L_1 = L_2 = 10.
\]

The partitioning matrices, analogous to base classical codes in HGP codes, are optimized to remove cycles using the Gradient Descent (GRADE) - Algorithmic Optimization (AO) method.
After GRADE-AO optimization, the resulting code eliminates cycles-4 and cycles-6 entirely and reduces the number of cycles-8 to 380.

The resulting 2D-SC-HGP code encodes $1624$ logical qubits into $5800$ physical qubits, yielding a rate of $0.28$.
The value 1624 exceeds the minimum $K=1600$ predicted earlier due to the presence of linearly dependent checks.

Ref.~\cite{yang2023spatially} also constructs a \( [[7300,2500]] \) SC-HGP code.
However, due to the difficulty in simulating the smaller \( [[5800,1624]] \) code at the circuit level, we did not explore simulating the larger one.
We leave this to future work, along with developing a full family of SC codes.

\section{Numerical Simulations}

To estimate the threshold error rate, we use the critical exponent ansatz~\cite{wang2003confinement}, which approximates the logical error rate as:

\begin{equation*}
\bar{P} = A + Bx + Cx^2,
\end{equation*} where  \( x = (p - p_{\text{th}}) d^{\alpha} \), \( p_{\text{th}} \) is the threshold error rate, \( d \) is the code distance, and \( \alpha \) is a critical exponent.

For the results shown in the main text, the fitted parameters for the HGP code family are:

\begin{equation*}
\begin{aligned}
    A &= 0.058 \pm 0.002,  &\quad B &= 0.10 \pm 0.02,  &\quad C &= 0.05 \pm 0.02, \\
    p_{\text{th}} &= 0.108 \pm 0.001,  &\quad \alpha &= 1.6 \pm 0.1.
\end{aligned}
\end{equation*}

For the LP code family, the fitted parameters are:

\begin{equation*}
\begin{aligned}
    A &= 0.037 \pm 0.001,  &\quad B &= 0.32 \pm 0.03,  &\quad C &= 0.7 \pm 0.2, \\
    p_{\text{th}} &= 0.103 \pm 0.001,  &\quad \alpha &= 0.67 \pm 0.04.
\end{aligned}
\end{equation*}

\end{document}